\documentclass{article}
\usepackage{spconf,amsmath,graphicx}
\usepackage{MnSymbol}
\usepackage{algorithmicx}
\usepackage{algpseudocode}
\usepackage{pgfplots}

\title{Approximate Least Squares\footnote{Preprint of the paper submitted to IEEE International Conference on Acoustics, Speech, and Signal Processing (ICASSP) 2014}}
\name{Michael Lunglmayr$^{\dagger}$ , Christoph Unterrieder$^{\dagger}$, Mario Huemer$^{\star}$}

\address{$^{\dagger}$ Klagenfurt University, Embedded Systems and Signal Processing, 9020 Klagenfurt, Austria \\
    $^{\star}$ Johannes Kepler University Linz, Institute of Signal Processing, 4040 Linz, Austria\\
    michael.lunglmayr@aau.at}

\begin{document}
\onecolumn
\maketitle
\sloppy
\begin{abstract}
We present a novel iterative algorithm for approximating the linear least squares 
solution with low complexity. 
After a motivation of the algorithm we discuss the algorithm's
properties including its complexity, and we present theoretical results as well as 
simulation based performance results.
We describe the analysis of its convergence behavior and show that in the noise free 
case the algorithm converges to the least squares solution.
\end{abstract}

\section{Introduction}
\label{sec:intro}
The linear least squares (LS) approach is an important and extensively
studied problem in many areas of signal processing with many 
practical applications from localization \cite{localization} 
to battery state estimation \cite{batterystate}. 
In applying the linear LS approach for a vector parameter $\bf x$,
we assume a signal model $\bf Hx$ disturbed by noise $\bf n$ such 
that
\begin{equation}
{\bf y} = {\bf Hx}  + {\bf n},
\label{eqn:systemmodel}
\end{equation}
where ${\bf H}$ is a known $m \times p$ observation matrix ($m \geq p$) with 
full rank $p$, ${\bf y}$ is a known $m \times 1$ vector 
(typically from measurements), ${\bf x}$ is an unknown $p \times 1$ parameter 
vector that is to be estimated and ${\bf n}$ is an $m \times 1$ noise vector.
For the LS approach, the statistical properties of $\bf n$ need not to be known.
For simplicity we only consider real vectors and matrices in this work, 
however, the presented concepts can  
easily be extended for complex vectors and matrices. 
The vector $\hat{\bf x}_{LS}$ that minimizes the cost function
\begin{align}
J(\hat{\bf x}) = \sum_{i=1}^m (y_i - {\bf h}_i^T \hat{\bf x})^2 = ( {\bf y} - {\bf H} \hat{\bf x} ) ^T ( {\bf y} - {\bf H} \hat{\bf x} )  
\end{align}
is the solution to the LS problem. Here 
${\bf h}_i^T$ is the $i^{th}$ row of $\bf H$ and 
$y_i$ is the $i^{th}$ element of $\bf y$, respectively. The LS 
solution is given by
\begin{equation}
\hat{\bf x}_{LS} = {\bf H}^\dag{\bf y},
\label{eqn:batch}
\end{equation}
with ${\bf H}^\dag = ({\bf H}^T{\bf H})^{-1}{\bf H}^T$ as the pseudoinverse 
of ${\bf H}$. Numerically more stable algorithms avoiding explicitly calculating ${\bf H}^\dag$, e.g. based on 
the QR decomposition, can for example be found in \cite{numeric}. 
A solution as in (\ref{eqn:batch}) is often called batch solution in literature \cite{kay}.

For real time applications one usually wants to avoid the calculation of the batch solution due to 
its computational complexity and its large memory requirements. Alternatives are
sequential algorithms such as the Sequential Least Squares (SLS) algorithm 
-- described in the next section -- or
gradient based approaches such as the iterative LS (ILS) \cite{numeric} algorithm. 
The latter algorithm is based on the steepest descent approach and iteratively calculates
\begin{align}
\hat{\bf x}^{(k)} = \hat{\bf x}^{(k-1)} - \mu \boldsymbol{\nabla}J(\hat{\bf x}^{(k-1)}),
\label{eqn:ls_iteration}
\end{align}
for iteration $k$. Here $\boldsymbol{\nabla}J(\hat{\bf x}^{(k-1)}) = - 2 {\bf H}^T {\bf y} +  2{\bf H}^T {\bf H} \hat{\bf x}^{(k-1)}$ 
is the gradient of $J(\hat{\bf x})$ at $\hat{\bf x}^{(k-1)}$.
For $k \rightarrow \infty$, $\hat{\bf x}^{(k)}$ converges to $\hat{\bf x}_{LS}$ given that
the iteration step width $\mu$ fulfills $0<\mu<1/(2 s_1^2({\bf H}))$ \cite{numeric}, 
with $s_1({\bf H})$ as the largest singular value of $\bf H$.
Alternatively, (\ref{eqn:ls_iteration}) can be written as
\begin{align}
\hat{\bf x}^{(k)} 
              &= \hat{\bf x}^{(k-1)} + \mu \sum_{i=1}^{m} 2 {\bf h}_{i}( {y_{i}} - {\bf h}_{i}^T \hat{\bf x}^{(k-1)}). \label{eqn:ILS_grad}
\end{align}
Analyzing the complexity of this approach one can see that $2pm+p$ multiplications are required per iteration.
In addition, every iteration of ILS requires the availability of all elements 
of the measurement vector $\bf y$.

Based on the principle of ILS we propose a novel iterative way of 
approximating 
the least squares solution that we call approximate least squares (ALS). 
As we will 
show, the complexity of this approach is significantly 
lower than for ILS and it requires only
one measurement value $y_i$ per iteration.

\begin{figure}[h]
\centering{\includegraphics[width=.5\columnwidth]{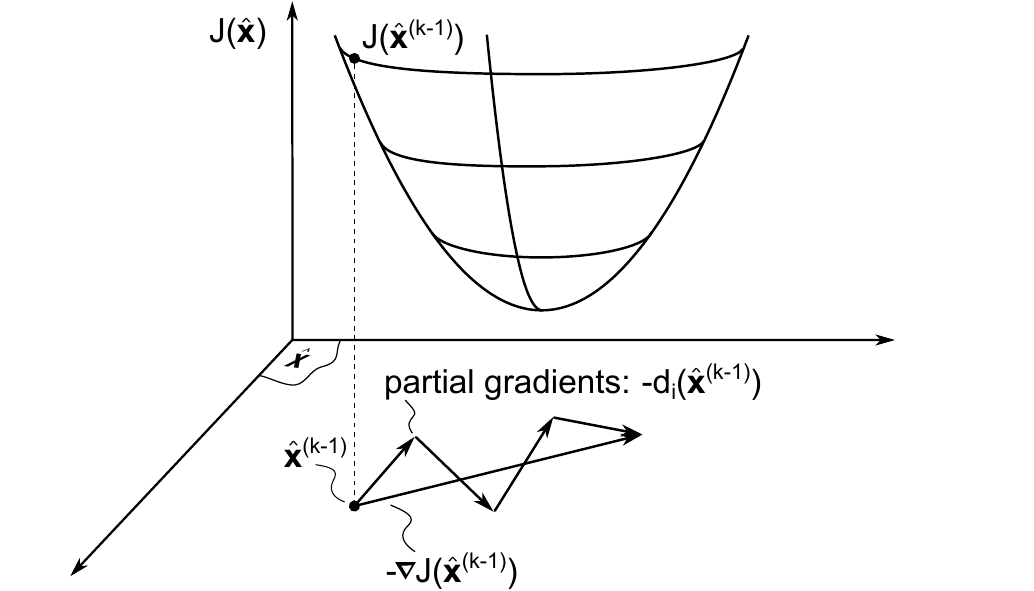}}
\caption{Gradient and partial gradients of ILS.\label{fig:partgrad}}
\end{figure}

When analyzing (\ref{eqn:ILS_grad}), the gradient can be interpreted as a 
sum of the partial gradients 
\begin{align}
{d}_i(\hat{\bf x}^{(k-1)}) = -2{\bf h}_{i}( {y_{i}} - {\bf h}_{i}^T \hat{\bf x}^{(k-1)})
\end{align}
as schematically depicted in Fig.~\ref{fig:partgrad}.
The idea of ALS is to use only \emph{one} of these partial gradients per iteration. 
Instead of moving a small step (due to $\mu$) in a steepest descent way in the negative direction of the gradient as done by ILS, 
ALS moves a small step in the negative direction of only a partial
gradient.
This has the advantage
of a lower complexity, but -- as we will discuss below -- also has the disadvantage of a higher noise sensitivity. 
Following this general idea,
two issues have to be addressed. First, the number of iterations 
of the algorithm to achieve satisfying performance results may be higher than the number of rows of $\bf H$. Second, the noise sensitivity has
to be reduced.
To cope with the first issue we suggest to re-use the rows ${\bf h}_i^T$ of {\bf H} in a cyclic manner.
Let the operator ``$\,\,\,\urcorner$'' be defined such that
for a positive natural number $i$: $i\urcorner = ( (i-1)\text{ mod } $m$ ) + 1$. From this it follows 
that $i\urcorner \in \{1,\ldots,m\}$. For better readability we do not
write the dependence of this operator 
on $m$ in the operator's symbol. For ALS, $m$ is always the number of rows of the matrix $\bf H$. 
An ALS iteration is now defined as 
\begin{align}
\hat{\bf x}^{(k)} &= \hat{\bf x}^{(k-1)} + \mu 2 {\bf h}_{k\urcorner} ({y_{k\urcorner}} - {\bf h}_{k\urcorner}^T \hat{\bf x}^{(k-1)}).
\label{eqn:als_main_equation}
\end{align}
That means for ALS that if $k$ reaches $m$, for the following iterations the first rows of $\bf H$ and the first 
elements of $\bf y$ are used again in a cyclic manner. 
We will now address the second issue, namely the noise sensitivity. 
As we will discuss below, if ${\bf n} = {\bf 0}$ then $\hat{\bf x}^{(k)}$ 
converges to $\hat{\bf x}_{LS}$ as $k \rightarrow \infty$. For the usual case
${\bf n} \neq {\bf 0}$, a noise dependent error remains. This
error can be greatly reduced by introducing a simple averaging 
process in the last $m$ iterations. A formal 
justification for this averaging will be given within the error analysis 
in Sect.~\ref{sec:conv_behav}. 
Summarizing, this leads to an overall formulation of the algorithm: \\ 
\clearpage
\vskip-6.5\smallskipamount
\noindent \rule{\columnwidth}{.4pt}
\vskip-1\smallskipamount
\noindent Algorithm: ALS\\
\vskip-7\smallskipamount
\noindent \rule{\columnwidth}{.4pt}
\begin{algorithmic}                    
\State $\hat{\bf x}_{ALS} = {\bf 0}$
\State $\hat{\bf x}^{(0)} = {\bf 0}$
\For {$k=1\ldots N$}
\State $\hat{\bf x}^{(k)} = \hat{\bf x}^{(k-1)} + \mu 2 {\bf h}_{k\urcorner} ({y_{k\urcorner}} - {\bf h}_{k\urcorner}^T \hat{\bf x}^{(k-1)})$
\If {$k > N-m$}
\State $\hat{\bf x}_{ALS} = \hat{\bf x}_{ALS} + \hat{\bf x}^{(k)}$
\EndIf
\EndFor
\State $\hat{\bf x}_{ALS} = \frac{1}{m} \hat{\bf x}_{ALS}$
\end{algorithmic}
\vskip-2\smallskipamount
\noindent \rule{\columnwidth}{.4pt}
Here $N$ denotes the number of iterations of the algorithm and 
$\hat{\bf x}_{ALS}$ is the approximation of $\hat{\bf x}_{LS}$ 
that is output by the algorithm.
When analyzing the above algorithm, and only counting the multiplications, one can 
see that $(2p+1)N$ overall multiplications are required to perform the algorithm. 
Compared to ILS a factor of around $m$ fewer multiplications per 
iteration are required. 
Although more iterations are usually needed for ALS  
its overall complexity is significantly lower
as will be demonstrated in Sect.~\ref{sec:sim_res}.
This decrease in complexity is bought with only a 
small degradation in performance. An additional advantage of ALS is 
that per iteration only one value $y_i$ and only one row 
${\bf h}_i^T$ of $\bf H$ are required. 
This significantly 
reduces the required number of other operations (additions,
memory accesses,...) and also simplifies the memory management as well 
as the architecture when thinking of a hardware implementation.
\section{Relation to prior work}  
\label{sec:prior} 
ALS not only has similarities to ILS but also to the SLS approach \cite{kay}.
For SLS the update equation
\begin{align}
\hat{\bf x}^{(k)} = \hat{\bf x}^{(k-1)} + {\bf K}_{k} ({y_{k}} - {\bf h}_{k}^T \hat{\bf x}^{(k-1)})
\end{align}
is sequentially calculated $m$ times, requiring an update of the gain vector 
${\bf K}_{k}$ at every iteration.
Although, the algorithm can deliver $\hat{\bf x}_{LS}$ after $m$ iterations, 
the update 
of ${\bf K}_{k}$ requires significant effort, including the multiplication of full 
matrices (although symmetry can be exploited
to reduce the complexity). ALS uses the same update equation 
(usually more than $m$ times),
with the simplified choice ${\bf K}_{k} = 2 \mu {\bf h}_{k\urcorner}$.

Update equation (8) is arithmetically similar to the Least Mean Squares 
(LMS) filter update step \cite{LMS}. However, the LMS update step uses a random 
(filter input) vector and one sample of a desired signal as input, whereas the ALS 
update step only uses one sample $y_i$ of the measurement vector $\bf y$ as input. 
Also the original formulation of the LMS algorithm for the so-called 
ADALINE \cite{LMS_first, ADALINE}
approach was based on a random input vector, providing an adaptive 
approach with a potentially unlimited set of input patterns.
Instead 
of the random input vector in the LMS case the deterministic and fixed rows of the 
observation matrix H are used in the update equation of the ALS. The row vectors 
${\bf h}_i^T$ and the measurement values $y_i$ are cyclically re-used. 
Another difference is 
the averaging at the last $m$ iterations which is unique for the ALS algorithm. And 
finally, the convergence behavior of ALS can be described in a completely 
deterministic manner, whereas the convergence of the LMS is usually only described 
in the mean. Anyhow, the authors are confident that some ideas improving LMS -- e.g. adjusting the 
step size \cite{LMS_variable, LMS_variable2} 
might be also
used to further improve the performance of ALS. 
\section{Convergence Behavior}
\label{sec:conv_behav} 
By rewriting (\ref{eqn:als_main_equation}) as
\begin{align}
  \hat{\bf x}^{(k)} &= ( {\bf I} - 2 \mu {\bf h}_{k\urcorner}
  {\bf h}_{k\urcorner}^T) \hat{\bf x}^{(k-1)} + 2 \mu {\bf h}_{k\urcorner} 
  {y_{k\urcorner}}
\end{align}
and defining the error vector of ALS 
${\bf e}^{(k)} =\hat{\bf x}^{(k)} - {\bf x}$ together with 
${\bf M}_{k\urcorner}= ( {\bf I} - 2 \mu {\bf h}_{k\urcorner} 
{\bf h}^T_{k\urcorner})$ one gets
\begin{align}
  {\bf x}^{(k)} &= ( {\bf I} - 2 \mu {\bf h}_{k\urcorner} 
  {\bf h}^T_{k\urcorner}) ({\bf x} + {\bf e}^{(k-1)})  
  + 2 \mu {\bf h}_{k\urcorner} {y_{k\urcorner}}\\
  &= {\bf x} - 2 \mu {\bf h}_{k\urcorner} {\bf h}_{k\urcorner}^T 
  {\bf x} + {\bf M}_{k\urcorner} {\bf e}^{(k-1)} +  2 \mu {\bf h}_{k\urcorner} {({\bf h}_{k\urcorner}^T{\bf x}+ n_{k\urcorner})}.
\end{align}
Subtracting $\bf x$ left and right from the equation leads to 
\begin{align}
{\bf e}^{(k)} = {\bf M}_{k\urcorner} {\bf e}^{(k-1)} +  
2 \mu {\bf h}_{k\urcorner} { n_{k\urcorner}}.
\label{eqn:first_error_splitting}
\end{align}
When defining 
${\bf \Delta}_{k\urcorner} = 2 \mu {\bf h}_{k\urcorner} n_{k\urcorner}$
one can write the above equation as 
\begin{align}
{\bf e}^{(k)} = &\prod_{i=1}^{k} {\bf M}_{i\urcorner} {\bf e}^{(0)} + \nonumber \\
  + &{\bf \Delta}_{k\urcorner} + {\bf M}_{(k-1)\urcorner}( {\bf \Delta}_{(k-2)\urcorner} + 
		  \ldots+ ( {\bf M}_{2}{\bf \Delta}_1) \ldots ), 
\end{align}
with ${\bf e}^{(0)}$ as the initial error.
Here 
the product of the matrices is defined as $\prod_{i=1}^{k} {\bf M}_{i\urcorner} = {\bf M}_{k\urcorner}{\bf M}_{(k-1)\urcorner} \ldots {\bf M}_{1}$.
When analyzing the above equation one can see that the error at iteration $k$ depends on the initial error ${\bf e}^{(0)}$
represented in ${\bf e}^{(k)}$ by the part ${\bf e}_0^{(k)}=\prod_{i=1}^{k} {\bf M}_{i\urcorner} {\bf e}^{(0)}$
as well as on an error term introduced by noise represented by ${\bf e}_\Delta^{(k)} = {\bf \Delta}_{k\urcorner} + {\bf M}_{(k-1)\urcorner}( {\bf \Delta}_{(k-2)\urcorner} + 
		  \ldots+ ( {\bf M}_{2}{\bf \Delta}_1) \ldots )$. With this one can write
\begin{align}
{\bf e}^{(k)} = {\bf e}_0^{(k)} + {\bf e}_\Delta^{(k)}.
\end{align}
If no noise is present then
\begin{align}
{\bf e}^{(k)} = {\bf e}_0^{(k)} = \prod_{i=1}^{k} {\bf M}_{i\urcorner} {\bf e}^{(0)}.
\end{align}
When choosing $k$ as an integer multiple of $m$ and defining 
${\bf M} = \prod_{i=1}^{m} {\bf M}_{i}$ one obtains
\begin{align}
{\bf e}^{(k)} = &\prod_{i=1}^{k} {\bf M}_{i\urcorner} {\bf e}^{(0)} = {\bf M}^{\frac{k}{m}} {\bf e}^{(0)}.
\end{align}
In \cite{journal} we show that for the choice
\begin{align}
0< \mu < \frac{1}{2\; \underset{i=1 \ldots m}{\text{max }}{\|{\bf h}^T_{i}\|_2^2}},
\end{align}
the matrix ${\bf M}$ has a $2$-norm smaller than one (although the proof is 
not complicated it is omitted here due to length constraints).  This implies that all  
eigenvalues of ${\bf M}$ have absolute values
smaller than one. 
From this it directly follows that ${\bf e}_0^{(k)}$ converges to zero as 
$k \rightarrow \infty$, i.e. $\hat{\bf x}_{ALS}=\hat{\bf x}_{LS}={\bf x}$.
This means that if no noise is present $\hat{\bf x}^{(k)}$ 
converges to ${\bf x}$.
However if ${\bf n} \neq {\bf 0}$ a persistent error ${\bf e}_\Delta^{(k)}$ remains.
In \cite{journal} we will give a more detailed analysis of
${\bf e}_\Delta^{(k)}$, showing that ${\bf e}^{(k)}$ features almost a 
periodic behavior 
from an index $k_p$ on, wherefrom ${\bf e}_0^{(k)}$ can be considered negligible.
This particular index $k_p$, which can also be specified analytically, can be 
used to define $N$ e.g. as $N=k_p+m$.

By analyzing the ALS algorithm one can see the importance of the averaging 
in the final $m$ iterations. 
As we already noted ${\bf e}^{(k)}$ is highly dependent on the noise for large $k$ (${\bf e}_0^{(k)}$
vanishes with increasing $k$). 
The averaging over the last $\hat{\bf x}^{(k)}$ vectors yields

\begin{align}
{\bf e}_{ALS} =  \hat{\bf x}_{ALS} - {\bf x} &= \left( \frac{1}{m} \sum_{k = N-m+1}^N \hat{\bf x}^{(k)} \right ) - {\bf x}\\
                                         &=  \frac{1}{m} \left(\sum_{k = N-m+1}^N \hat{\bf x}^{(k)}  - m{\bf x}\right )\\
                                         &=  \frac{1}{m} \sum_{k = N-m+1}^N \left(\hat{\bf x}^{(k)}  - {\bf x}\right )\\
                                         &=  \frac{1}{m} \sum_{k = N-m+1}^N {\bf e}^{(k)}.
\end{align}
That means by averaging over the last $m$ vectors $\hat{\bf x}^{(k)}$ an averaging 
over the 
corresponding error vectors occurs.
Since for a practical application it is highly unlikely that all these error vectors 
have equal length and 
point in the same direction
(in this case averaging would have no effect) this averaging step typically 
significantly reduces the error norm.
The averaging only has to be done \emph{once}, it therefore presents only a minor 
complexity 
increase (overall only $pm$ additions and $p$ multiplications with the constant 
$1/m$).

\section{Simulation Results} \label{sec:sim_res} 
\begin{figure}[tb]
\begin{center}
\includegraphics[width=.6\columnwidth]{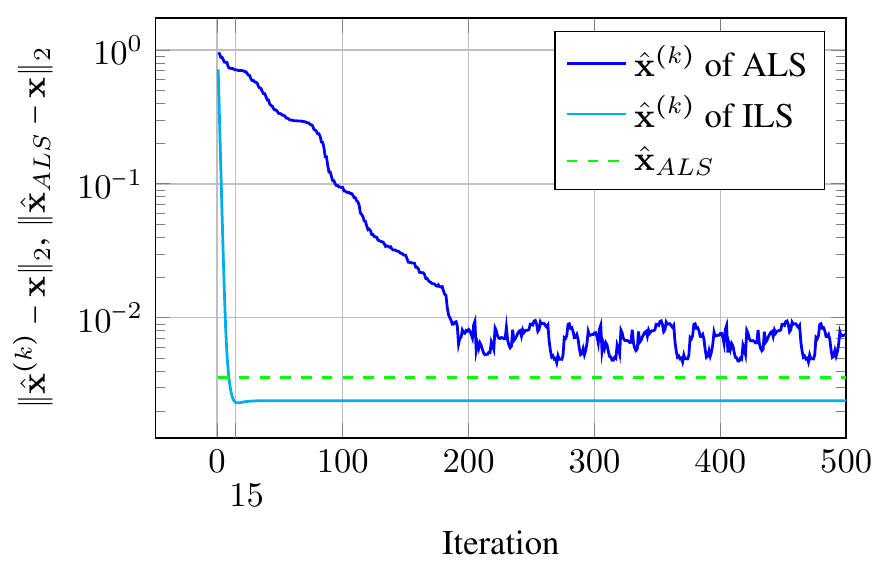}
\caption{ Example of errors of ILS and ALS \label{fig:error_sin} }
\end{center}
\end{figure}

\begin{figure}[tb]
\begin{center}
\includegraphics[width=.6\columnwidth]{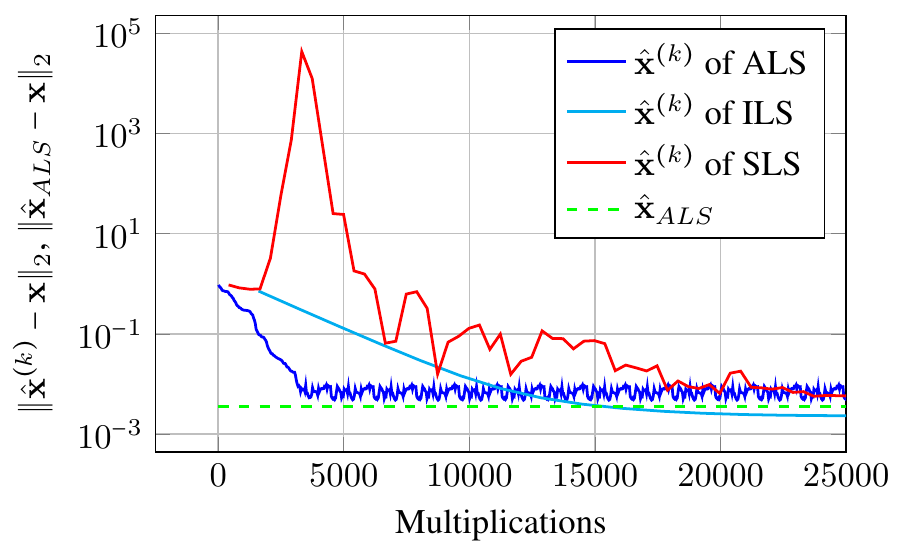}
\caption{ Error norms over to the number of multiplications \label{fig:error_sin_mul} }
\end{center}
\end{figure}

We first show simulation results of a typical example of least squares estimation: the estimation
of amplitudes of sine signals in noise. For this demonstration example we 
chose ${\bf H}$ as a $100 \times 8$ matrix 
with elements $H_{n,k} = \text{cos}(2 \pi n T_s f_k)$.
The frequencies $f_k$ are not necessarily integer multiples of a base frequency.
The elements of the noise vector have been 
sampled indepentently from a normal distribution with zero mean and a 
standard deviation $\sigma = 10^{-2}$. 
The amplitudes $\bf x$ 
have been estimated using $100$ values, forming the vector $\bf y$. 
The step size for ILS was chosen as $\mu=1/(2.05\,s_1^2({\bf H}))$ and
for ALS as $\mu=1/(2.05\,\text{max}_{i=1 \ldots m}{\|{\bf h}^T_{i}\|_2^2})$.
The estimation performance
has been measured by calculating the norm of the difference vector 
between the true vector $\bf x$ and the estimated vectors, respectively. Fig.~\ref{fig:error_sin}
shows a typical simulation result for ILS and ALS. 
In this figure, $\hat{\bf x}_{ALS}$ is the estimated parameter vector resulting 
after averaging the 
final $m$ out of $N$ vectors $\hat{\bf x}^{(k)}$, represented as a horizontal line for 
illustration purposes.
As one can see, ILS requires significantly less
iterations than ALS, but with about $m=100$ times more multiplications per iteration. 
The
performance of ALS is only slightly worse than the performance of ILS
but ALS features a significantly lower overall complexity. 
In this figure one can observe an interesting behavior of ALS. After a certain number of iterations 
the influence of ${\bf e}_0^{(k)}$ becomes negligible. 
This reflects in an oscillatory behavior 
of the error norm as can be seen in Fig.~\ref{fig:error_sin}.
This oscillatory behavior comes from the fact that the values $y_i$ are cyclically 
re-used in the $N$ ALS iterations. As a consequence also the noise values appear
in a cyclic manner.
The averaging at the end of ALS is most effective if $N$ is chosen large enough
so that the effects of ${\bf e}_0^{(k)}$ are negligible. Such a value for $N$ can be found with 
simulations or based on analytical results as will be presented in \cite{journal}.
To provide a fair comparison, in Fig.~\ref{fig:error_sin_mul} we compared ALS, ILS and SLS in 
terms of its error norms over the number of
calculated multiplications. 
As one can see, if the error performance of 
ALS is sufficient for a given application, its complexity is significantly lower. 
In this example ILS needs about $3$ times more multiplications than ALS to obtain
the same error norm. 
But as stated above,
this complexity analysis is only based on the number of multiplications per iteration. 
Including other operations (additions, memory accesses) would furthermore favor 
ALS. Due to page constraints we omitted a more detailed complexity analysis in this paper. 
If the error performance of ALS is not sufficient for 
a given application one could choose a different approach, but extended variants of the ALS, e.g. with 
adjusting $\mu$ during the iterations
show promising first results towards 
further reducing the error norm. One can immediately see the benefits of such an approach in
(\ref{eqn:first_error_splitting}) because 
the noise dependent part of the error vector scales with $\mu$, as will be
described in detail in \cite{journal}. 
But as extensive performance simulations showed, ALS' performance is on average very close to the LS solution. 
Table.~\ref{tab:perf_res} shows performance results for random $\bf H$ matrices. 
The entries of these matrices have been sampled from a uniform distribution out of $[0,1]$. Every 
simulation has been done for white Gaussian noise with 
$\sigma \in S=\{ 10^{-5},10^{-4},10^{-3},10^{-2},10^{-1}, 1\}$, 
respectively,
with $100$ random matrices $\bf H$ per $\sigma$ value and $100$ 
random vectors $\bf x$
(with random entries also sampled from a uniform
distribution out of $[0,1]$) per $\bf H$ matrix. For every $\sigma$ 
value the averages $\overline{||\hat{\bf x}_{ALS}-{\bf x}||_2}$ and 
$\overline{||\hat{\bf x}_{LS}-{\bf x}||_2}$
over the simulated results have been calculated.
Table.~\ref{tab:perf_res} shows the maximum relative increase of ALS'
averaged error norm over the averaged error norms of LS, whereas the maximization
has been done over the elements of $S$: 
$r_{max} = \text{max}_S \left(1-\frac{\overline{||\hat{\bf x}_{ALS}-{\bf x}||_2}}{\overline{||\hat{\bf x}_{LS}-{\bf x}||_2}} \right)$. 
We furthermore want to note that 
the relative increase of the averaged
error norms remained nearly constant over all simulated $\sigma$ values.
As one can see in this table, the performance of ALS shows on average only a minor degradation 
compared to the LS solution.
\begin{table}[htb]
\centering
\begin{tabular}{|l|c||l|c|}
\hline
dim$({\bf H})$ & $r_{max}$ & dim$({\bf H})$ & $r_{max}$ \\
\hline
$100  \times 1$   & $9.3\%$  & $1000 \times 1$   & $9.5\%$ \\
$100  \times 2$   & $9.7\%$  & $1000 \times 2$   & $9.5\%$ \\
$100  \times 3$   & $11.2\%$ & $1000 \times 3$   & $10.7\%$\\
$100  \times 5$   & $11.3\%$ & $1000 \times 5$   & $12.8\%$\\
$100  \times 10$  & $16\%$   & $1000 \times 10$  & $15.3\%$\\    
\hline
\end{tabular}
\caption{Performance results for random matrices. \label{tab:perf_res} }
\end{table}
\section{Conclusion} \label{sec:sim_res}
We presented a novel algorithm for 
approximating the solution of the linear least squares problem. We discussed its 
convergence behavior and
demonstrated that the algorithm provides a close solution to the least squares solution
with low complexity. The presented algorithm 
shows promising potential for further extension in theory and implementation as well 
for use in a variety of applications.

\vfill\pagebreak

\end{document}